\documentclass[aps,onecolumn,showpacs,nofootinbib]{revtex4}
\usepackage{epsf,amsfonts,amssymb,amsbsy,rotate}

\def\slashchar#1{\setbox0=\hbox{$#1$}           
   \dimen0=\wd0                                 
   \setbox1=\hbox{/} \dimen1=\wd1               
   \ifdim\dimen0>\dimen1                        
      \rlap{\hbox to \dimen0{\hfil/\hfil}}      
      #1                                        
   \else                                        
      \rlap{\hbox to \dimen1{\hfil$#1$\hfil}}   
      /                                         
   \fi}                                         %
\begin{document}
\title{{\large\sc Gold-plated Mode of {\bf CP}-Violation in Decays of
$\boldsymbol B_{c}$ Meson from QCD Sum Rules}}
\author{V.V.Kiselev}
\email{kiselev@th1.ihep.su} \affiliation{Russian State Research
Center ``Institute for High Energy Physics'', Protvino, Moscow
Region, 142281 Russia \\ Fax: 7-0967-744937}
\pacs{13.25.Gv,12.15.Hh,11.30.Er,11.55.Hx}

\begin{abstract}
  The model-independent method based on the triangle ideology is
  implemented to extract the CKM-matrix angle $\gamma$ in the decays
  of doubly heavy long-lived meson $B_c$. We analyze a color structure of
  diagrams and conditions to reconstruct two reference-triangles
  by tagging the flavor and CP eigenstates of
  $D^0\leftrightarrow \bar D^0$ mesons in the fixed
  exclusive channels. The characteristic branching ratios are
  evaluated in the framework of QCD sum rules.
\end{abstract}
\maketitle

\section{Introduction}
\label{sec:1}

The $B_c$ meson first observed by the CDF collaboration at FNAL
\cite{cdf} is expected to be copiously produced in the future
experiments at hadron colliders \cite{revbc} with facilities
oriented to the study of fine effects in the heavy quark
interactions such as the parameters of CP-violation and charged
weak current mixing\footnote{See, for instance, the program on the
B physics at Tevatron \cite{BRunII}.}. So, one could investigate
the spectroscopy, production mechanism and decay features of $B_c$
\cite{Bj,OPEBc,QCDSRBc} with the incoming sample of several
billion events. In such circumstances, in addition to the current
success in the experimental study of decays with the CP-violation
in the gold-plated mode of neutral $B$-meson by the BaBar and
Belle collaborations \cite{BB} allowing one to extract the
CKM-matrix angle $\beta$ in the unitarity triangle, a possible
challenge is whether one could get an opportunity to extract some
information about the CKM unitarity triangle from the $B_c$
physics in a model independent way or not. The theoretical
principal answer is one can do it. Indeed, there is an intriguing
opportunity to extract the angle $\gamma$ in the model-independent
way using the strategy of reference triangles \cite{Gronau} in the
decays of doubly heavy hadrons. This ideology for the study of
CP-violation in $B_c$ decays was originally offered by M.Masetti
\cite{Masetti}, independently investigated by R.Fleischer and
D.Wyler \cite{FW} and extended to the case of doubly heavy
baryons\footnote{A review on the physics of doubly heavy baryons
is given in \cite{revqq}.} in \cite{QQCP}.

Let us point out necessary conditions to extract the CP-violation
effects in the model-independent way.
\begin{enumerate}
\item Interference. The measured quantities have to involve the
amplitudes including both the CP-odd and CP-even phases. \item
  Exclusive channels. The hadronic final state has to be fixed in
  order to isolate a definite flavor contents and, hence, the
  definite matrix elements of CKM matrix, which can exclude the
  interference of two CP-odd phases with indefinite CP-even phases due
  to strong interactions at both levels of the quark structure and the
  interactions in the final state.
\item Oscillations. The definite involvement of the CP-even phase
is
  ensured by the oscillations taking place in the systems of
  neutral $B$ or $D$ mesons, wherein the CP-breaking effects can be
  systematically implemented.
\item Tagging. Once the oscillations are involved, the tagging of
both
  the flavor and CP eigenstates is necessary for the complete
  procedure.
\end{enumerate}
The gold-plated modes in the decays of neutral $B$ mesons involve
the oscillations of mesons themselves and, hence, they require the
time-dependent measurements. In contrast, the decays of doubly
heavy hadrons such as the $B_c$ meson and $\Xi_{bc}$ baryons with
the neutral $D^0$ or $\bar D^0$ meson in the final state do not
require the time-dependent measurements. The triangle ideology is
based on the direct determination of absolute values for the set
of four decays, at least: the decays of hadron in the tagged $D^0$
meson, the tagged $\bar D^0$ meson, the tagged CP-even
state\footnote{The CP-odd states of $D^0$ can be used, too.
However, their registration requires the detection of CP-even
state of $K^0$, which can be complicated because of a detector
construction, say, by a long base of $K^0$ decay beyond a tracking
system.} of $D^0$, and the decay of the anti-hadron into the
tagged CP-even state of $D^0$. To illustrate, let us consider the
decays of
$$
B^+_c\to D^0 D_s^+, \quad\mbox{and}\quad B^+_c\to \bar D^0 D_s^+.
$$

\begin{figure*}[th]
  \begin{center}
\setlength{\unitlength}{1.3mm} \hspace*{2cm}
\begin{picture}(90,25)
\put(-10,0){\epsfxsize=55\unitlength \epsfbox{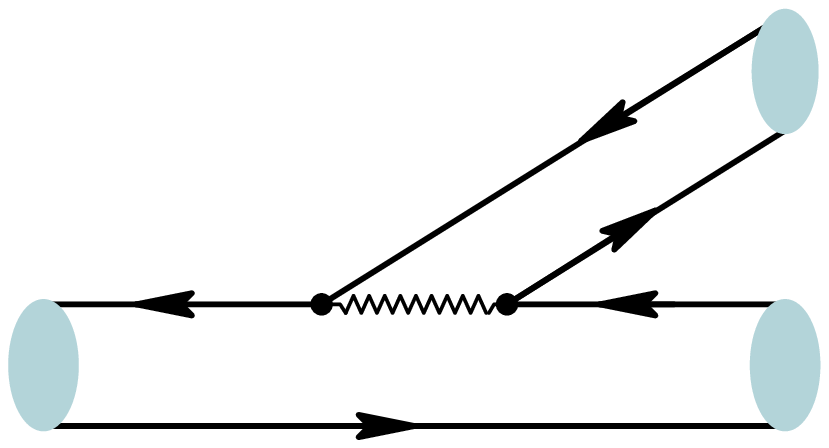}}
\put(33,0){\epsfxsize=55\unitlength \epsfbox{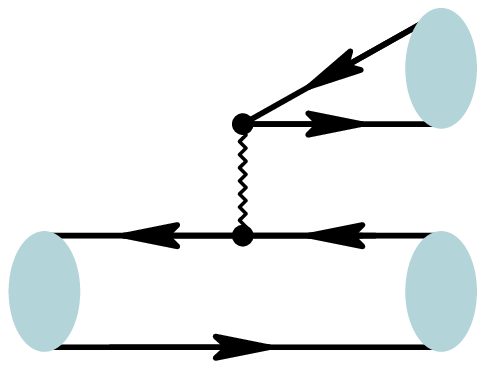}}
\put(-2.3,6.5){$B^+_{c}$} \put(34.,6.5){$D^+_{s}$}
\put(34.2,21){$D^0$} \put(47.2,6.55){$B^+_{c}$}
\put(67.5,6.6){$D^0$} \put(67.5,18.1){$D^+_{s}$}
\put(5.,11.5){$\scriptstyle b$} \put(15.,2.){$\scriptstyle c$}
\put(26.,20.8){$\scriptstyle u$} \put(27,15.){$\scriptstyle c$}
\put(29,11.5){$\scriptstyle s$} \put(53.5,12){$\scriptstyle b$}
\put(63.5,11.5){$\scriptstyle u$} \put(57.5,2){$\scriptstyle c$}
\put(63.2,20.2){$\scriptstyle s$} \put(63,14.2){$\scriptstyle c$}
\end{picture}
    \caption{The diagrams of $\bar b$-quark decay contributing to the weak
     transition $B^+_c\to D^0 D_s^+$.}
    \label{fig:1}
  \end{center}
\end{figure*}
\begin{figure*}[th]
  \begin{center}
\setlength{\unitlength}{1.3mm} \hspace*{0cm}
\begin{picture}(90,25)
\put(-10,2){\epsfxsize=55\unitlength \epsfbox{1.eps}}
\put(42,1){\epsfxsize=55\unitlength \epsfbox{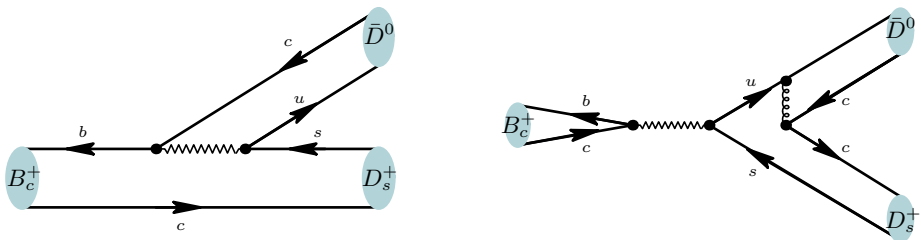}}
\put(-2.3,8.5){$B^+_{c}$} \put(34.,8.5){$D^+_{s}$}
\put(34.2,23){$\bar D^0$} \put(48.2,13.9){$B^+_{c}$}
\put(87.5,4.2){$D^+_{s}$} \put(87.5,23.6){$\bar D^0$}
\put(5.,13.5){$\scriptstyle b$} \put(15.,4.){$\scriptstyle c$}
\put(26.,22.8){$\scriptstyle c$} \put(27,17.){$\scriptstyle u$}
\put(29,13.5){$\scriptstyle s$} \put(56.5,16.7){$\scriptstyle b$}
\put(73.5,9.5){$\scriptstyle s$} \put(56.5,12){$\scriptstyle c$}
\put(73.2,18.7){$\scriptstyle u$} \put(83,16.8){$\scriptstyle c$}
\put(83,12.){$\scriptstyle c$}
\end{picture}
    \caption{The diagrams of $\bar b$-quark decay contributing to the weak
     transition $B^+_c\to \bar D^0 D_s^+$.}
    \label{fig:1a}
  \end{center}
\end{figure*}

The corresponding diagrams with the decay of $\bar b$-quark are
shown in Figs. \ref{fig:1} and \ref{fig:1a}. We stress that two
diagrams of the decay to $D^0$ have the additional negative sign
caused by the Pauli interference of two charmed quarks, which,
however, completely compensated after the Fierz transformation for
the corresponding Dirac matrices.

The exclusive modes make the penguin terms to be excluded, since
the penguins add an even number of charmed quarks, i.e. two or
zero, while the final state contains two charmed quarks including
one from the $\bar b$ decay and one from the initial state.
However, the diagram with the weak annihilation of two
constituents, i.e. the charmed quark and beauty anti-quark in the
$B_c^+$ meson, can contribute in the next order in $\alpha_s$ as
shown in Fig. \ref{fig:1a} for the given final state.
Nevertheless, we see that such the diagrams have the same
weak-interaction structure as at the tree level. Therefore, they
do not break the consideration under interest. The magnitude of
$\alpha_s$-correction to the absolute values of corresponding
decay widths is discussed in Section \ref{sec:2}.

\begin{figure}[th]
  \begin{center}
\setlength{\unitlength}{1mm}
\begin{picture}(80,50)
\put(10,7){\epsfxsize=75\unitlength \epsfbox{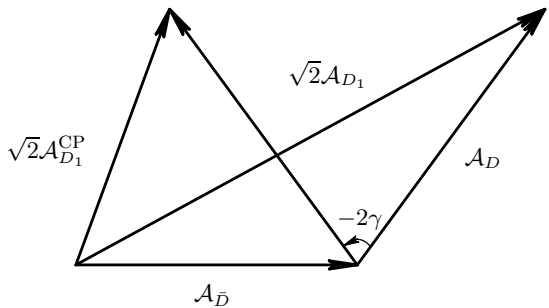}}
\put(32,4){${\cal A}_{\bar D}$} \put(68,22){${\cal A}_{D}$}
\put(51.,14){$-2\gamma$} \put(44.5,33){$\sqrt{2}{\cal A}_{D_1}$}
\put(7,23){$\sqrt{2}{\cal A}_{D_1}^{\mbox{\sc cp}}$}
\end{picture}
    \caption{The reference-triangles.}
    \label{fig:3}
  \end{center}
\end{figure}

Thus, the CP-odd phases of decays under consideration are
determined by the tree-level diagrams shown in Figs. \ref{fig:1}
and \ref{fig:1a}. Therefore, we can write down the amplitudes in
the following form:
\begin{equation}
  \label{eq:3}
  {\cal A}(B^+_c\to D^0 D_s^+) \stackrel{\mbox{\tiny def}}{=}
  {\cal A}_{D} = V^*_{ub} V_{cs}\cdot {\cal M}_{D},\qquad
  {\cal A}(B^+_c\to \bar D^0 D_s^+) \stackrel{\mbox{\tiny def}}{=}
  {\cal A}_{\bar D} = V^*_{cb} V_{us}\cdot {\cal M}_{\bar D},
\end{equation}
where ${\cal M}_{\bar D,\,D}$ denote the CP-even factors depending on
the dynamics of strong interactions. Using the definition of angle
$\gamma$
$$
\gamma \stackrel{\mbox{\tiny def}}{=} -{\rm arg}\left[\frac{V_{ub} V^*_{cs}}
{V_{cb} V^*_{us}}\right],
$$
for the CP-conjugated channels\footnote{For the sake of simplicity
  we put the overall phase of arg $V_{cb} V^*_{us}=0$, which
  corresponds to fixing the representation of the CKM matrix, e.g. by
  the Wolfenstein form \cite{Wolf}.} we find
\begin{equation}
  \label{eq:4}
  {\cal A}(B^-_c\to \bar D^0 D_s^-)= e^{-2{\rm i}\gamma}
  {\cal A}_{D}, \qquad
  {\cal A}(B^-_c\to D^0 D_s^-)=  {\cal A}_{\bar D}.
\end{equation}
We see that the corresponding widths for the decays to the flavor
tagged modes coincide with the CP-conjugated ones. However, the story
can be continued by using the definition of CP-eigenstates for the
oscillating $D^0\leftrightarrow \bar D^0$ system\footnote{The
  suppressed effects of CP-violation in the oscillations of neutral
  $D$ mesons are irrelevant here, and we can neglect them in the sound
  way.},
$$
D_{1,\,2} =\frac{1}{\sqrt{2}}(D^0\pm \bar D^0),
$$
so that we straightforwardly get
\begin{eqnarray}
  \label{eq:5}
 \sqrt{2}{\cal A}(B^+_c\to D_s^+ D_1)
 \stackrel{\mbox{\tiny def}}{=}
 \sqrt{2} {\cal A}_{D_1} &=& {\cal A}_{D}+{\cal A}_{\bar D},\\[2mm]
 \sqrt{2}{\cal A}(B^-_c\to D_s^- D_1)
 \stackrel{\mbox{\tiny def}}{=}
 \sqrt{2} {\cal A}_{D_1}^{\mbox{\sc cp}}& = &e^{-2{\rm i}\gamma}
 {\cal A}_{D}+{\cal A}_{\bar D}.
\label{eq:6}
\end{eqnarray}
The complex numbers entering (\ref{eq:5}) and (\ref{eq:6})
establish two triangles with the definite angle $2\gamma$ between
the vertex positions as shown in Fig. \ref{fig:3}.
Thus, due to the unitarity, the measurement of four absolute values
\begin{eqnarray}
  \label{eq:7}
  |{\cal A}_{D}| = |{\cal A}(B^+_c\to D_s^+ D^0)|,
&\quad &
  |{\cal A}_{\bar D}| = |{\cal A}(B^+_c\to D_s^+ \bar D^0)|, \nonumber\\[2mm]
  |{\cal A}_{D_1}| = |{\cal A}(B^+_c\to D_s^+ D_1)|,
&\quad &
  |{\cal A}_{D_1}^{\mbox{\sc cp}}| = |{\cal A}(B^-_c\to D_s^- D_1)|,
\end{eqnarray}
can constructively reproduce the angle $\gamma$ in the model-independent way.

The above triangle-ideology can be implemented for the analogous
decays to the excited states of charmed mesons in the final state.

The residual theoretical challenge is to evaluate the
characteristic widths or branching fractions. We address this
problem and analyze the color structure of amplitudes. So, we find
that the matrix elements under interest have the different
magnitudes of color suppression, so that at the tree level we get
${\cal A}_{D}\sim O(\sqrt{N_c})$ and ${\cal A}_{\bar D}\sim
O(1/\sqrt{N_c})$, while the ratio of relevant CKM-matrix elements,
$$
\left|\frac{V_{ub} V^*_{cs}}{V_{cb} V^*_{us}}\right|\sim O(1)
$$
with respect to the small parameter of Cabibbo angle,
$\lambda = \sin\theta_C$, which one can easily find in the Wolfenstein
parametrization
$$
V_{\mbox{\sc ckm}} = \left(\begin{array}{ccc}
V_{ud} & V_{us} & V_{ub} \\[2mm]
V_{cd} & V_{cs} & V_{cb} \\[2mm]
V_{td} & V_{ts} & V_{tb}
\end{array}\right)
=
\left(\begin{array}{ccc}
1-\frac{1}{2}\lambda^2 & \lambda & A \lambda^3 (\rho-{\rm i}\eta) \\[2mm]
-\lambda & 1-\frac{1}{2}\lambda^2  & A \lambda^2 \\[2mm]
A \lambda^3 (1-\rho-{\rm i}\eta) & -A \lambda^2 & 1
\end{array}\right).
$$
Nevertheless, the interference of two diagrams in the decays of
$B_c^+$ to the $D^0$ meson is destructive, and the absolute values
of the amplitudes ${\cal A}_{D}$ and ${\cal A}_{\bar D}$ become
close to each other. Thus, we expect that the sides of the
reference-triangles are of the same order of magnitude, which
makes the method to be an attractive way to extract the angle
$\gamma$.

In Section \ref{sec:2} we classify the diagrams for the decays of
doubly heavy meson $B_c^+$ by the color and weak-interaction
structures. Section \ref{sec:3} is devoted to the numerical
estimates in the framework of QCD sum rules. The results are
summarized in Conclusion.

\section{Color structures}
\label{sec:2}

In the framework of $1/N_c$-expansion we have got the following
scaling rules of color structures in the processes with the
hadrons composed of the quark and anti-quark:
\begin{enumerate}
\item
The meson wavefunction
$$
\Psi_M \sim\frac{1}{\sqrt{N_c}}\,{\delta^i}_j.
$$
\item The coupling constant
$$
\alpha_s \sim \frac{1}{{N_c}}.
$$
\item
The Casimir operators
$$
C_A=N_c,\qquad C_F=\frac{N_c^2-1}{2N_c}\sim O(N_c).
$$
\item The Fierz relation for the generators of SU($N_c$) group in
the fundamental representation
$$
{t^{A\,i}}_j\, {t^{A\,k}}_m =\frac{C_F}{N_c}\, {\delta^i}_m\,{\delta^k}_j
- \frac{1}{N_c}\, {t^{A\,i}}_m\,{t^{A\,k}}_j.
$$
\end{enumerate}

Next, the non-leptonic weak Lagrangian has a form  typically
given by the following term \cite{BBL}:
\begin{equation}
  \label{eq:7a}
  {\cal H}_{\rm eff} = \frac{G_F}{2\sqrt{2}}\,
V_{cb}(\bar b^i\Gamma_\mu c_j)\,
V^*_{us}(\bar u^k\Gamma^\mu s_l)\,
C_{\pm}\left({\delta^i}_j\,{\delta^k}_l\pm{\delta^i}_l\,{\delta^k}_j\right)
+\ldots
\end{equation}
where $\Gamma_\mu =\gamma_\mu(1-\gamma_5)$, and the Wilson coefficients
$$
C_{\pm}\sim O(1)
$$
in the $1/N_c$-expansion.

Then, we can proceed with the analysis of decays under interest.


Two diagrams shown in Fig. \ref{fig:1} scale with the different
order in $1/N_c$, so that
$$
{\cal A}_{1D} \sim  \frac{1}{\sqrt{N_c}},\qquad {\cal A}_{2D} \sim
{\sqrt{N_c}}.
$$
More definitely we get the color factors
\begin{equation}
  \label{eq:8}
  {\cal F}^c_{1D} =\sqrt{N_c}\,a_2, \qquad
  {\cal F}^c_{2D} =\sqrt{N_c}\,a_1,
\end{equation}
where
\begin{eqnarray}
  \label{eq:9}
  a_1 &=& \frac{1}{2N_c}\,\left[C_+(N_c+1)+C_-(N_c-1)\right],\\[1mm]
  a_2 &=& \frac{1}{2N_c}\,\left[C_+(N_c+1)-C_-(N_c-1)\right].
\end{eqnarray}
Nevertheless, we have to calculate both diagrams in order to take
into account their interference which is not suppressed by any
kinematical factors except the color factor of $1/N_c$.

Two diagrams shown in Fig. \ref{fig:1a} also scale with the
different order in $1/N_c$, so that
$$
{\cal A}_{1\bar D} \sim  \frac{1}{\sqrt{N_c}},\qquad {\cal
A}_{2\bar D} \sim {\sqrt{N_c}}.
$$
More definitely we get the color factors
\begin{equation}
  \label{eq:8a}
  {\cal F}^c_{1\bar D} =\sqrt{N_c}\,a_2, \qquad
  {\cal F}^c_{2\bar D} =\sqrt{N_c}\,a_1\,C_F\, 4\pi\alpha_s ,
\end{equation}
where we have explicitly included the $\alpha_s$ correction.
However, we can easily find that due to the virtualities of quarks
and gluons the kinematical suppression of second diagram in Fig.
\ref{fig:1a} is given by the factor of
$$
{\cal S}\sim |\tilde \Psi(0)|^2\,\frac{\alpha_s(k_g^2)}{k_g^2}\,
\frac{N_c}{\Delta E_q},
$$
where
$$
|\tilde\Psi(0)|^2 \sim \Lambda^3_{\mbox{\sc qcd}}
$$
is the characteristic value of wavefunction, the gluon virtuality
is determined by the expression
$$
k_g^2 = m_c^2 (v_2+v_D)^2 =
m_c^2\,\frac{M_1^2-(M_2-M_3)^2}{M_2M_3} \approx
{(m_b+m_c)^2}+O(m_{s,\,d}/m_{c,\,b})\gg \Lambda^2_{\mbox{\sc
qcd}},
$$
with $M_{1,2,3}$ being the masses of the mesons in the initial and
final states, respectively, and the virtuality of quark line
connected to the virtual gluon is of the order of
$$
\Delta E_q \sim m_{c}.
$$
Therefore,
$$
{\cal S} \sim \frac{\Lambda^3_{\mbox{\sc qcd}}}{m_{c,\,b}^3}\,
\frac{1}{\ln m_{c,\,b}/\Lambda_{\mbox{\sc qcd}}}\ll 1.
$$
Therefore, the above $\alpha_s$-corrections can be neglected to
the leading order in the $1/m_Q$-expansion.

Finally, in this section we have analyzed the color and
weak-interaction structures of decay amplitudes and isolate those
of the largest magnitude, while the numerical estimates are
presented in the next section.

\section{Numerical estimates}
\label{sec:3}

In this section we use the machinery of QCD sum rules
\cite{QCDSR,SR3pt} in order to calculate the widths and branching
ratios for the gold-plated modes under study.

The calculations of exclusive non-leptonic modes usually involves
the approximation of factorization \cite{fact}, which, as
expected, can be quite accurate for the $B_c$, since the
quark-gluon sea is suppressed in the heavy quarkonium. Thus, the
important parameters are the factors $a_1$ and $a_2$ in the
non-leptonic weak lagrangian, which depend on the normalization
point suitable for the $B_c$ decays. In this way, we, first,
calculate the form factors in the semileptonic transitions
\cite{KT,KLO,KKL} and, second, evaluate the non-leptonic matrix
elements in the factorization approach.

We accept the following convention on the normalization of wave
functions for the hadron states under study, i.e. for the
pseudoscalar ($P$) and vector ($V$) states:
\begin{eqnarray}
&& \langle 0|{\cal J}_\mu|P\rangle =  -{\rm i}\, f_P\,p_\mu,\label{pseudo}\\[1mm]
&& \langle 0|{\cal J}_\mu|V\rangle =  \epsilon_\mu\, f_V\,M_V,
\end{eqnarray}
where $f_{P,\,V}$ denote the leptonic constants, so that they are
positive,
$$
f_{P,\,V} > 0,
$$
$p_\mu$ is a four-momentum of the hadron, $\epsilon_\mu$ is a
polarization vector of $V$, $M_V$ is its mass, and the current is
composed of the valence quark fields constituting the hadron
$$
{\cal J}_\mu = \bar q_1\, \gamma_\mu(1-\gamma_5)\,q_2.
$$
In this respect we can easily apply the ordinary Feynman rules for
the calculations of diagrams, so that the quark-meson vertices in
the decay channel are chosen with the following spin structures:
$$
\Gamma_P = \frac{\rm i}{\sqrt{2}}\,\gamma_5,\qquad \Gamma_V = -
\frac{1}{\sqrt{2}}\,\epsilon_\mu.
$$
Then, we get general expressions for the hadronic matrix elements
of weak currents in the exclusive decays of $P\to P^\prime$ and
$P\to V$ with the definitions of form factors given by the
formulae
\begin{eqnarray}
\langle P^\prime(p_2)|{\cal J}_\mu |P(p_1)\rangle &=& f_{+}
p_{\mu} +
f_{-}q_{\mu},\\
\frac{1}{\rm i}\langle V(p_2)|{\cal J}_\mu|P(p_1)\rangle &=& {\rm
i} F_V\epsilon_{\mu\nu\alpha\beta}\epsilon^{*\nu}
p^{\alpha}q^{\beta}+
F_0^A\epsilon_{\mu}^{*} + 
F_{+}^{A}(\epsilon^{*}
p_1)p_{\mu} + F_{-}^{A}(\epsilon^{*} p_1)q_{\mu},
\end{eqnarray}
where $q_{\mu} = (p_1 - p_2)_{\mu}$ and $p_{\mu} = (p_1 +
p_2)_{\mu}$. The form factors $f_{\pm}$ are dimensionless, while
$F_V$ and $F^A_{\pm}$ has a dimension of inverse energy, $F^A_0$
is of the energy dimension. In the case of nonrelativistic
description for both initial and final meson states we expect that
$$
\begin{array}{rrr}
f_+ > 0, & f_- < 0, & F_V > 0, \\[2mm]
F^A_0 > 0, & F^A_+ < 0, & F^A_- > 0.
\end{array}
$$
It is important to note that for the pseudoscalar state the
hermitian conjugation in (\ref{pseudo}) does not lead to the
change of sign in the right hand side of equation because of the
prescription accepted, since the conjugation of imaginary unit
takes place with the change of sign for the momentum of meson (the
transition from the out-state to in-one). The same speculations
show that the spin structure of matrix element in the quark-loop
order does not involve a functional dependence of form factors on
the transfer momentum squared except of $F^A_0$, so that we expect
that the simplest modelling in the form of the pole dependence can
be essentially broken for $F^A_0$, while the other form factors
are fitted by the pole model in a reasonable way.

Following the standard procedure for the evaluation of form
factors in the framework of QCD sum rules \cite{SR3pt}, in the
$B_c$ decays we consider the three-point functions
\begin{widetext}
\begin{eqnarray}
&& \Pi_{\mu}(p_1, p_2, q^2) = {\rm i}^2 \int {\rm d}x\,{\rm
d}y\,e^{{\rm
i}(p_2\cdot x - p_1\cdot y)} \cdot 
\langle 0|T\{\bar q_2(x)\gamma_5 q_1(x)\, {\cal J}_\mu(0)\, \bar
b(y)\gamma_5
c(y)\}|0 \rangle,\\
&& \Pi_{\mu\nu}^{{\cal J}}(p_1, p_2, q^2) = {\rm i}^2 \int {\rm
d}x\,{\rm
d}y\,e^{{\rm i}(p_2\cdot x - p_1\cdot y)} \cdot 
\langle 0|T\{\bar q_2(x)\gamma_{\mu} q_1(x)\, {\cal J}_\mu(0)\,
\bar b(y)\gamma_5 c(y)\}|0\rangle,
\end{eqnarray}
\end{widetext}
where $\bar q_2(x)\gamma_5 q_1(x)$ and $\bar
q_2(x)\gamma_{\nu}q_1(x)$ denote the interpolating currents for
the final states mesons.

The standard procedure for the evaluation of correlators and form
factors is described in \cite{KT,KLO,KKL,exBc}. The most important
notes are the following \cite{KT,KLO,KKL,exBc}:
\begin{itemize}
    \item
    For the heavy quarkonium $\bar b c$, where the relative velocity of
    quark movement is small, an essential role is taken by the
    Coulomb-like $\alpha_s/v$-corrections.
    \item
    We have found that the normalizations of leptonic constants for the heavy
    quarkonia are fixed by the appropriate choice of effective constant for
    the coulomb exchange $\alpha_s^{\cal C}$, while the stability is very
    sensitive to the prescribed value of heavy quark mass. Thus,
    these parameters of sum rules are extracted from the two-point
    QCD sum rules with a quite good accuracy.
    \item
    In the framework of two-point sum rules for the heavy-light channel
    the fixed value of threshold energy $E_c$ determines the binding energy
    of heavy quark in the meson, $\bar \Lambda\approx 0.63$ GeV,
    which yields the same value of mass for the beauty quark,
    $m_b\approx 4.6$ GeV, as it was determined from the analysis of two-point
    sum rules in the $\bar b b$ channel. However, taking into account
    the second order corrections in $1/m_c$, we find that the mass
    of charmed quark is shifted to the value of $m_c\approx 1.2$ GeV
    in the heavy-light channel in comparison with the $\bar c c$
    states. Thus, in the transition of $B_c$ to the charmed meson we put
    the mass of spectator charmed quark equal to $m_c\approx 1.2$
    GeV.
    \item
    In the framework of the effective theory of heavy quarks with the
    expansion in the inverse masses of the heavy quarks
    \cite{HQET,NRQCD,pNRQCD,vNRQCD}, the spin symmetry
    relations between the form factors in the soft limit of zero recoil momentum
    can be derived \cite{KKL}.
    \item
    We take the following ordinary ratios of leptonic constants for
    the vector and pseudoscalar states and for the heavy-strange
    mesons:
    $$ \frac{f_{B^*}}{f_B} \approx \frac{f_{D^*}}{f_D} \approx 1.11,\qquad
    \frac{f_{B_s}}{f_B} \approx \frac{f_{D_s}}{f_D} \approx 1.16,
    $$
    which agree with both the lattice computations \cite{latt} and the
    estimates in the framework of potential models taking into account
    relativistic corrections \cite{Faust'}.
    \item
    The leptonic constant of $B_c$ is taken from a scaling relation for the heavy
    quarkonia \cite{scale}.
\end{itemize}

\begin{table*}[th]
\caption{The form factors of various transitions calculated in the
framework of QCD sum rules at $q^2=0$ in comparison with the
estimates in the potential model (PM) of \cite{PMK}.} \label{form}
\begin{tabular}{|c|c|c|c|c|c|c|}
\hline Transition & $f_+$, [PM] & $f_-$, [PM] & $F_V$, [PM]
(GeV$^{-1}$) & $F_0^A$,
[PM] (GeV) & $F_+^A$, [PM] (GeV$^{-1}$) & $F_-^A$, [PM] (GeV$^{-1}$) \\
\hline $B_c\to D^{(*)}$ & 0.32, [0.29] & -0.34, [-0.37] & 0.20,
[0.21] & 3.6, [3.6] &
-0.062, [-0.060] & 0.10, [0.16] \\
$B_c\to D_s^{(*)}$ & 0.45, [0.43] & -0.43, [-0.56] & 0.24, [0.27]
& 4.7, [4.7] &
-0.077, [-0.071] & 0.13, [0.20] \\
\hline
\end{tabular}
\end{table*}
Our estimates are summarized in Table \ref{form} extracted from
\cite{exBc}, where for the sake of comparison we expose the
results obtained in the potential model \cite{PMK}, which
parameters are listed in Appendix B of ref.\cite{KKL}. In the
potential model the most reliable results are expected at zero
recoil of meson in the final state of transition, since the wave
functions are rather accurately calculable at small virtualities
of quarks composing the meson. We take the predictions of the
potential model at zero recoil and evolve the values of form
factors to zero transfer squared in the model with the pole
dependence
$$
F_i(q^2) = \frac{F_i(0)}{1-q^2/{M_{i,\,{\rm pole}}^2}},
$$
making use of numerical values of $M_{i,\,{\rm pole}}$ shown in
Table \ref{pole}. We stress the fact that the potential model
points to the approximately constant value of the form factor
$F_0^A$ because of additional kinematical dependence in the
transition of $B_c\to D^*$ and $B_c\to D_s^*$.

\begin{table*}[th]
\caption{The pole masses used in the model for the form factors in
various transitions.} \label{pole}
\begin{tabular}{|l|c|c|c|c|c|c|}
\hline Transition & $M_{\rm pole}[f_+]$, GeV & $M_{\rm
pole}[f_-]$, GeV & $M_{\rm pole}[F_V]$, GeV & $M_{\rm
pole}[F_0^A]$, GeV & $M_{\rm pole}[F_+^A]$, GeV &
$M_{\rm pole}[F_-^A]$, GeV \\
\hline
$B_c\to D^{(*)}$ & 5.0 & 5.0 & 6.2 & $\infty$ & 6.2 & 6.2 \\
$B_c\to D_s^{(*)}$ & 5.0 & 5.0 & 6.2 & $\infty$ & 6.2 & 6.2 \\
\hline
\end{tabular}
\end{table*}

Next, we investigate the validity of spin-symmetry relations in
the $B_c$ decays to $D^{(*)}$ and $D_s^{(*)}$. The results of
estimates for the $f_{\pm}$ evaluated by the symmetry relations
with the inputs given by the form factors $F_V$ and $F_0^A$
extracted from the sum rules are presented in Table \ref{symm} in
comparison with the values calculated in the framework of sum
rules.

\begin{table*}[th]
\caption{The comparison of sum rule results for the form factors
$f_{\pm}$ with the values obtained by the spin symmetry with the
inputs of $F_0^A$ and $F_V$ extracted from the QCD sum rules. The
values of pole masses are also explicitly shown.} \label{symm}
\begin{tabular}{|l|c|c|c|c|c|}
\hline Transition & Form factor & Sum Rules & Spin symmetry &
$M_{\rm pole}[f_{\pm}]$,
GeV & $M_{\rm pole}[F_V]$, GeV \\
\hline
$B_c\to D^{(*)}$ & $f_+$ & 0.32 & 0.31 & 4.8 & 6.2 \\
$B_c\to D^{(*)}$ & $f_-$ & -0.34 & -0.36 & 4.8 & 6.2 \\
$B_c\to D_s^{(*)}$ & $f_+$ & 0.45 & 0.45 & 4.8 & 6.2 \\
$B_c\to D_s^{(*)}$ & $f_-$ & -0.43 & -0.51 & 4.8 & 6.2 \\
\hline
\end{tabular}
\end{table*}
\begin{table}[h!tb]
\caption{Exclusive non-leptonic decay widths of the $B_c$ meson,
$\Gamma$ in $10^{-15}$ GeV. The $\bar b$-quark decays with
$c$-quark spectator. }
\begin{center}
\begin{tabular}{|c|l|r|r|r|r|r|r|}
\hline Class & ~~~~~Mode & $\Gamma$ \cite{exBc} & $\Gamma$
\cite{vary} & $\Gamma$ \cite{chch} & $\Gamma$ \cite{narod}
& $\Gamma$ \cite{IKP} & $\Gamma$ \cite{CdF}\\
\hline & $B_c^+ \rightarrow D^+ \overline
D^{\hspace{1pt}\raisebox{-1pt}{$\scriptscriptstyle 0$}}$ &$1.9
\,a_2^2$ &$0.633 \,a_2^2 $ & $0.664  \,a_2^2 $ & $2.72  \,a_2^2 $
& $0.753  \,a_2^2 $ & $0.15  \,a_2^2 $\\
& $B_c^+ \rightarrow D^+ \overline
D^{\hspace{1pt}\raisebox{-1pt}{$\scriptscriptstyle *0$}}$ &$2.75
\,a_2^2 $ &$0.762 \,a_2^2 $ & $0.695  \,a_2^2 $ & $2.10  \,a_2^2 $
& $1.925  \,a_2^2 $ & $0.13  \,a_2^2 $\\
& $B_c^+ \rightarrow  D^{\scriptscriptstyle *+} \overline
D^{\hspace{1pt}\raisebox{-1pt}{$\scriptscriptstyle 0$}}$ &$1.8
\,a_2^2 $ &$0.289 \,a_2^2 $ & $0.653  \,a_2^2 $ & $0.86  \,a_2^2 $
& $0.399  \,a_2^2 $ & $1.46  \,a_2^2 $\\
& $B_c^+ \rightarrow  D^{\scriptscriptstyle *+} \overline
D^{\hspace{1pt}\raisebox{-1pt}{$\scriptscriptstyle *0$}}$ &$12.0
\,a_2^2 $ &$0.854 \,a_2^2 $ & $1.080  \,a_2^2 $ & $1.32  \,a_2^2 $
& $1.95  \,a_2^2 $ & $2.4  \,a_2^2 $\\
II & $B_c^+ \rightarrow D_s^+ \overline
D^{\hspace{1pt}\raisebox{-1pt}{$\scriptscriptstyle 0$}}$ &$0.18
\,a_2^2 $ &$0.0415 \,a_2^2 $& $0.0340 \,a_2^2 $ &
& $0.0405  \,a_2^2 $ & $0.01  \,a_2^2 $\\
& $B_c^+ \rightarrow D_s^+ \overline
D^{\hspace{1pt}\raisebox{-1pt}{$\scriptscriptstyle *0$}}$ &$0.25
\,a_2^2 $ &$0.0495 \,a_2^2 $& $0.0354 \,a_2^2 $ & & $0.101 \,a_2^2
$
& $0.009  \,a_2^2 $\\
& $B_c^+ \rightarrow  D_s^{\scriptscriptstyle *+} \overline
D^{\hspace{1pt}\raisebox{-1pt}{$\scriptscriptstyle 0$}}$ &$0.17
\,a_2^2 $ &$0.0201 \,a_2^2 $& $0.0334 \,a_2^2 $ & & $0.0222
\,a_2^2 $
& $0.087  \,a_2^2 $\\
& $B_c^+ \rightarrow  D_s^{\scriptscriptstyle *+} \overline
D^{\hspace{1pt}\raisebox{-1pt}{$\scriptscriptstyle *0$}}$ &$0.93
\,a_2^2 $ &$0.0597 \,a_2^2 $& $0.0564 \,a_2^2 $ & & $0.109 \,a_2^2
$
& $0.15  \,a_2^2 $\\
\hline
\end{tabular}
\label{I,II}
\end{center}
\end{table}

We have found that the uncertainty in the estimates is basically
determined by the variation of pole masses in the
$q^2$-dependencies of form factors, which govern the evolution
from the zero recoil point to the zero transfer squared. \noindent
So, the variation of $M_{\rm pole}[f_{\pm}]$ in the range of
$4.8-5$ GeV for the transitions of $B_c\to D^{(*)}$ and $B_c\to
D_s^{(*)}$ results in the 30\%-uncertainty in the form factors
presented in Table \ref{symm}.

The QCD SR estimates for the non-leptonic decays of $\bar b$-quark
in $B_c$ as obtained in \cite{exBc} and the present work give the
widths represented in Tables \ref{I,II} and \ref{III} in
comparison with the values calculated in the potential models. The
sum rule predictions are significantly greater than the estimates
of the potential models for the transitions with the color
permutation, i.e. for the class II processes\footnote{The class I
processes, which amplitudes are proportional to the factor of
$a_1$, without the color permutations in the effective lagrangian
are not involved in the modes under consideration.} with the
factor of $a_2$.

Further, for the transitions, wherein the interference is
significantly involved, the class III processes, we find that the
absolute values of different terms given by the squares of $a_1$
and $a_2$ calculated in the sum rules are greater than the
estimates of potential models. We stress that under fixing the
definitions of hadron state phases as described in the beginning
of section \ref{sec:3}, we have found that the Pauli interference
has determined the negative sign of two amplitudes with $a_1$ and
$a_2$, however, the relevant Fierz transformation has led to the
complete cancellation of the Pauli interference effect, and the
relative sign of two amplitudes in the modes under consideration
is positive in agreement with the results of potential models
listed in Table \ref{III}. Taking into account the negative value
of $a_2$ with respect to $a_1$, we see that all of decays shown in
Table \ref{III} should be suppressed in comparison with the case
of the interference switched off. The characteristic values of
effects caused by the interference is presented in Table
\ref{Pauli}, where we put the widths in the form
$$
\Gamma = \Gamma_0 + \Delta \Gamma,\qquad \Gamma_0=x_1\,
a_1^2+x_2\, a_2^2,\quad \Delta\Gamma =z a_1\, a_2.
$$
Then, we conclude that the interference can be straightforwardly
tested in the listed decays, wherein its significance reaches
about 40\%.

\begin{table*}[ht]
\caption{Exclusive non-leptonic decay widths of the $B_c$ meson,
$\Gamma$ in $10^{-15}$ GeV, the symbol $\star$ marks the result of
this work. The $\bar b$-quark decays involving the Pauli
interference with the $c$-quark spectator. }
\begin{center}
\begin{tabular}{|c|l|r|r|}
\hline Class & ~~~~~Mode & $\Gamma$ [$\star$]\hspace*{9mm} &
$\Gamma$ \cite{IKP} \hspace*{7mm}\\
\hline & $B_c^+ \rightarrow D^+
D^{\hspace{1pt}\raisebox{-1pt}{$\scriptscriptstyle 0$}}$ &
$(0.023\,a_1+0.023\,a_2)^2$ & $(0.0147\,a_1+0.0146\,a_2)^2$ \\
& $B_c^+ \rightarrow D^+
D^{\hspace{1pt}\raisebox{-1pt}{$\scriptscriptstyle *0$}}$ &
$(0.022\,a_1+0.025\,a_2)^2$ & $(0.0107\,a_1+0.0234\,a_2)^2$ \\
& $B_c^+ \rightarrow  D^{\scriptscriptstyle *+}
D^{\hspace{1pt}\raisebox{-1pt}{$\scriptscriptstyle 0$}}$ &
$(0.025\,a_1+0.022\,a_2)^2$ & $(0.0233\,a_1+0.0106\,a_2)^2$ \\
& $B_c^+ \rightarrow  D^{\scriptscriptstyle *+}
D^{\hspace{1pt}\raisebox{-1pt}{$\scriptscriptstyle *0$}}$ &
$(0.051\,a_1+0.051\,a_2)^2$ & $(0.0235\,a_1+0.0235\,a_2)^2$ \\
III & $B_c^+ \rightarrow D_s^+
D^{\hspace{1pt}\raisebox{-1pt}{$\scriptscriptstyle 0$}}$ &
$(0.11\,a_1+0.14\,a_2)^2$ & $(0.0689\,a_1+0.672\,a_2)^2$ \\
& $B_c^+ \rightarrow D_s^+
D^{\hspace{1pt}\raisebox{-1pt}{$\scriptscriptstyle *0$}}$
& $(0.11\,a_1+0.15\,a_2)^2$ & $(0.0503\,a_1+0.106\,a_2)^2$ \\
& $B_c^+ \rightarrow  D_s^{\scriptscriptstyle *+}
D^{\hspace{1pt}\raisebox{-1pt}{$\scriptscriptstyle 0$}}$
& $(0.12\,a_1+0.13\,a_2)^2$ & $(0.101\,a_1+0.0498\,a_2)^2$ \\
& $B_c^+ \rightarrow  D_s^{\scriptscriptstyle *+}
D^{\hspace{1pt}\raisebox{-1pt}{$\scriptscriptstyle *0$}}$
& \begin{tabular}{c} $0.067\,a_1^2+0.706\,a_2^2$\\[-1mm]
$+0.14\,a_1 a_2$\end{tabular} & $(0.104\,a_1+0.110\,a_2)^2$ \\
\hline
\end{tabular}
\label{III}
\end{center}
\end{table*}

\begin{table}[ht]
\caption{The effect of interference in the exclusive non-leptonic
decay widths of the $B_c$ meson with the $c$-quark as spectator at
$a_1^b =1.14$ and $a_2^b=-0.20$.}
\begin{center}
\begin{tabular}{|l|c|}
\hline
~~~~~Mode & $\Delta\Gamma/\Gamma_0$, \% \\
\hline $B_c^+ \rightarrow D^+
D^{\hspace{1pt}\raisebox{-1pt}{$\scriptscriptstyle 0$}}$ &
-34 \\
 $B_c^+ \rightarrow D^+
D^{\hspace{1pt}\raisebox{-1pt}{$\scriptscriptstyle *0$}}$ &
-38 \\
 $B_c^+ \rightarrow  D^{\scriptscriptstyle *+}
D^{\hspace{1pt}\raisebox{-1pt}{$\scriptscriptstyle 0$}}$ &
-30 \\
 $B_c^+ \rightarrow  D^{\scriptscriptstyle *+}
D^{\hspace{1pt}\raisebox{-1pt}{$\scriptscriptstyle *0$}}$ &
-34 \\
 $B_c^+ \rightarrow D_s^+
D^{\hspace{1pt}\raisebox{-1pt}{$\scriptscriptstyle 0$}}$ &
-43 \\
 $B_c^+ \rightarrow D_s^+
D^{\hspace{1pt}\raisebox{-1pt}{$\scriptscriptstyle *0$}}$
& -45 \\
 $B_c^+ \rightarrow  D_s^{\scriptscriptstyle *+}
D^{\hspace{1pt}\raisebox{-1pt}{$\scriptscriptstyle 0$}}$
& -37 \\
 $B_c^+ \rightarrow  D_s^{\scriptscriptstyle *+}
D^{\hspace{1pt}\raisebox{-1pt}{$\scriptscriptstyle *0$}}$
& -55 \\
\hline
\end{tabular}
\label{Pauli}
\end{center}
\end{table}
The predictions of QCD sum rules for the exclusive decays of $B_c$
are summarized in Table \ref{common} at the fixed values of
factors $a_{1,2}$ and lifetime. For the sake of completeness and
comparison we show the estimates for the channels with the neutral
$D$ meson and charged one $D^+$ as well as for the vector states
in addition to the pseudoscalar ones.

First, we see that the similar decay modes without the strange
quark in the final state can be, in principle, used for the same
extraction of CKM angle $\gamma$, however, this channels are more
problematic from the methodic point of view, since the sides of
reference-triangles significantly differ from each
other\footnote{The ratio of widths is basically determined by the
factor of $|V_{cb} V_{ud} a_2|^2/|V_{ub} V_{cd} a_1|^2\sim 110$,
if we ignore the interference effects.}, so that the measurements
have to be extremely accurate in order to get valuable information
on the angle. Indeed, we should accumulate a huge statistics for
the dominant mode in order to draw any conclusion on the
consistency of triangle with a small side.

Second, the decay modes with the vector neutral $D$ meson in the
final state are useless for the purpose of the CKM measurement
under the approach discussed. However, the modes with the vector
charged $D^*$ and $D_s^*$ mesons can be important for the
procedure of $\gamma$ extraction. This note could be essential for
the mode with $D^{*+}\to D^0 \pi^+$ and $D^0\to K^-\pi^+$, but, in
this case, the presence of neutral charmed meson should be
carefully treated in order to avoid the misidentification with the
primary neutral charmed meson. In other case, we should use the
mode with the neutral pion $D^{*+}\to D^+ \pi^0$, which detection
in an experimental facility could be problematic. The same note is
applicable for the vector $D_s^{*+}$ meson, which radiative
electromagnetic decay is problematic for the detection, too, since
the photon could be loosed. However, the lose of the photon for
the fully reconstructed $D_s^+$ and $B_c^+$ does not disturb the
analysis.

\begin{table*}[th]
\caption{Branching ratios of exclusive $B_c^+$ decays at the fixed
choice of factors: $a_1^b=1.14$ and $a_2^b=-0.20$ in the
non-leptonic decays of $\bar b$ quark. The lifetime of $B_c$ is
appropriately normalized by $\tau[B_c] \approx 0.45$ ps.}
\label{common}
\begin{center}
\begin{tabular}{|l|c|}
\hline
~~~~~Mode & BR, $10^{-6}$\\
\hline
 $B_c^+ \rightarrow D^+
\overline D^{\hspace{1pt}\raisebox{-1pt}{$\scriptscriptstyle 0$}}$
 & 53\\
 $B_c^+ \rightarrow D^+
\overline D^{\hspace{1pt}\raisebox{-1pt}{$\scriptscriptstyle
*0$}}$
 & 75\\
 $B_c^+ \rightarrow  D^{\scriptscriptstyle *+}
\overline D^{\hspace{1pt}\raisebox{-1pt}{$\scriptscriptstyle 0$}}$
 & 49\\
 $B_c^+ \rightarrow  D^{\scriptscriptstyle *+}
\overline D^{\hspace{1pt}\raisebox{-1pt}{$\scriptscriptstyle
*0$}}$
 & 330\\
 $B_c^+ \rightarrow D_s^+ \overline
D^{\hspace{1pt}\raisebox{-1pt}{$\scriptscriptstyle 0$}}$
 & 4.8\\
 $B_c^+ \rightarrow D_s^+
\overline D^{\hspace{1pt}\raisebox{-1pt}{$\scriptscriptstyle
*0$}}$
 & 7.1\\
 $B_c^+ \rightarrow  D_s^{\scriptscriptstyle *+} \overline
D^{\hspace{1pt}\raisebox{-1pt}{$\scriptscriptstyle 0$}}$
 & 4.5\\
 $B_c^+ \rightarrow  D_s^{\scriptscriptstyle *+}
\overline D^{\hspace{1pt}\raisebox{-1pt}{$\scriptscriptstyle
*0$}}$
 & 26\\
\hline
\end{tabular}
\begin{tabular}{|l|c|}
\hline
~~~~~Mode & BR, $10^{-6}$\\
\hline
 $B_c^+ \rightarrow D^+
D^{\hspace{1pt}\raisebox{-1pt}{$\scriptscriptstyle 0$}}$
 & 0.32\\
 $B_c^+ \rightarrow D^+
D^{\hspace{1pt}\raisebox{-1pt}{$\scriptscriptstyle *0$}}$
 & 0.28\\
 $B_c^+ \rightarrow  D^{\scriptscriptstyle *+}
D^{\hspace{1pt}\raisebox{-1pt}{$\scriptscriptstyle 0$}}$
 & 0.40\\
 $B_c^+ \rightarrow  D^{\scriptscriptstyle *+}
D^{\hspace{1pt}\raisebox{-1pt}{$\scriptscriptstyle *0$}}$
 & 1.59\\
 $B_c^+ \rightarrow D_s^+
D^{\hspace{1pt}\raisebox{-1pt}{$\scriptscriptstyle 0$}}$
 & 6.6\\
 $B_c^+ \rightarrow D_s^+
D^{\hspace{1pt}\raisebox{-1pt}{$\scriptscriptstyle *0$}}$
 & 6.3\\
 $B_c^+ \rightarrow  D_s^{\scriptscriptstyle *+}
D^{\hspace{1pt}\raisebox{-1pt}{$\scriptscriptstyle 0$}}$
 & 8.5\\
 $B_c^+ \rightarrow  D_s^{\scriptscriptstyle *+}
D^{\hspace{1pt}\raisebox{-1pt}{$\scriptscriptstyle *0$}}$
 & 40.4\\
\hline
\end{tabular}
\end{center}
\end{table*}

In the above estimates we put the following values of parameters:
\begin{itemize}
    \item the leptonic constants:
    $
    f_{D} = 0.22\,\mbox{GeV},\quad
    f_{D^*} = 0.24\,\mbox{GeV},\quad
    f_{D_s} = 0.24\,\mbox{GeV},\quad
    f_{D_s^*} = 0.27\,\mbox{GeV};
    $
    \item the CKM elements:
    $
    |V_{ub}| = 0.003,\quad
    |V_{cb}| = 0.04,\quad
    |V_{cs}| = 0.975,
    $
\end{itemize}
so that the numbers can be appropriately scaled at other values of
input parameters.

In the BTeV \cite{BRunII} and LHCb \cite{LHCB} experiments one
expects the $B_c$ production at the level of several billion
events. Therefore, we predict $10^4-10^5$ decays of $B_c$ in the
gold-plated modes under interest. The experimental challenge is
the efficiency of detection. One usually get a 10\%-efficiency for
the observation of distinct secondary vertices outstanding from
the primary vertex of beam interaction. Next, we have to take into
account the branching ratios of $D_s$ and $D^0$ mesons. This
efficiency crucially depends on whether we can detect the neutral
kaons and pions or not. So, for the $D_s$ meson the corresponding
branching ratios grow from 4\% (no neutral $K$ and $\pi$) to 25\%.
The same interval for the neutral $D^0$ is from 11 to 31\%. The
detection of neutral kaon is necessary for the measurement of
decay modes into the CP-odd state $D_2$ of the neutral $D^0$
meson, however, we can omit this cross-check channel from the
analysis dealing with the CP-even state of $D_1$. The
corresponding intervals of branching ratios reachable by the
experiment are from 0.5 to 1.3\% for the CP-even state and from
1.5 to 3.8\% for the CP-odd state of $D^0$. The pessimistic
estimate for the product of branching ratios is about $2\cdot
10^{-4}$, which results in $2-20$ reconstructed events. Thus, an
acceptance of experimental facility and an opportunity to detect
neutral pions and kaons as well as reliable estimates of total
cross section for the $B_c$ production in hadronic collisions are
of importance in order to make expectations more accurate.

\section{Conclusion}
\label{sec:con}

In this work we have shown how the reference-triangle ideology can
be used for the model-independent extraction of CKM-matrix angle
$\gamma$ from the set of branching ratios of doubly heavy meson
$B_c$ exclusively decaying to the neutral $D$ mesons. Tagging the
flavor and CP-eigenstates of such the $D$ mesons allows one to
avoid the uncertainties caused by the QCD dynamics of quarks.

We have estimated the characteristic branching ratios in the
framework of QCD sum rules, which yields the values of the order
of
$$
{\cal B}[B_c^+\to D_s^+ \bar D^0] \approx 5\cdot 10^{-6}.
$$
Accepting the above value, and putting the efficiency of tagging
procedure equal to 0.5\% for the neutral charmed meson and 4\% for
the charmed strange meson in the final state as well as the vertex
reconstruction efficiency equal to 10\%, we can expect the
observation of about 10 reconstructed events per year at the LHC
collider in such the experiment like LHCB or in BTeV experiment at
FNAL.

The author thanks Prof. R.Dzhelyadin for the possibility to visit
CERN in collaboration with the LHCB group, to which members I
express a gratitude for a hospitality. I am grateful to Profs.
A.K.Likhoded and M.A.Ivanov for clarifications and a cooperation,
to Prof. V.F.Obraztsov, Yu.Gouz, O.Yushchenko and V.Romanovsky for
useful discussions and valuable remarks.

This work is in part supported by the Russian Foundation for Basic
Research, grants 01-02-99315, 01-02-16585.



\begin{thebibliography}{**}
\bibitem{cdf}
F. Abe et al., CDF Collaboration, Phys. Rev. Lett. {\bf 81}, 2432
(1998),
Phys. Rev. {\bf D58}, 112004 (1998).
\bibitem{revbc}
S.S.Gershtein et al., Sov. J. Nucl. Phys. {bf 48}, 327 (1988),
Yad. Fiz. {\bf 48}, 515 (1988);\\
S.~S.~Gershtein, V.~V.~Kiselev, A.~K.~Likhoded and
A.~V.~Tkabladze,
Phys.\ Usp.\  {\bf 38}, 1 (1995) [Usp.\ Fiz.\ Nauk {\bf 165}, 3
(1995)]
[arXiv:hep-ph/9504319];\\
E.Eichten, C.Quigg, Phys. Rev. {\bf D49}, 5845 (1994);\\
S.S.Gershtein et al., Phys. Rev. {\bf D51}, 3613 (1995);\\
A.~V.~Berezhnoi, V.~V.~Kiselev, A.~K.~Likhoded and
A.~I.~Onishchenko,
Phys.\ Atom.\ Nucl.\  {\bf 60}, 1729 (1997) [Yad.\ Fiz.\  {\bf
60N10},  (1997)] [arXiv:hep-ph/9703341].
\bibitem{BRunII}
K.~Anikeev {\it et al.},
arXiv:hep-ph/0201071.
\bibitem{Bj}
J.D.Bjorken, draft report 07/22/86 (1986) [unpublished];\\
M.Lusignoli, M.Masetti, Z. Phys. {\bf C51}, 549 (1991);\\
V.V.Kiselev, A.K.Likhoded, A.V.Tkabladze, Phys. Atom. Nucl.  {\bf
56}, 643 (1993), Yad. Fiz. {\bf 56}, 128 (1993);\\
V.V.Kiselev, A.V.Tkabladze, Yad. Fiz. {\bf 48}, 536 (1988).
\bibitem{OPEBc}
I.Bigi, Phys. Lett. {\bf B371}, 105 (1996);\\
M.Beneke, G.Buchalla, {Phys. Rev.} {\bf D53}, 4991 (1996);\\
A.I.Onishchenko, [hep-ph/9912424];\\
Ch.-H.Chang, Sh.-L.Chen, T.-F.Feng, X.-Q.Li, Commun. Theor. Phys.
{\bf 35}, 51 (2001),
Phys. Rev. {\bf D64}, 014003 (2001).
\bibitem{QCDSRBc}
P.Colangelo, G.Nardulli, N.Paver, Z.Phys. {\bf C57}, 43 (1993);\\
E.Bagan et al., Z. Phys. {\bf C64}, 57 (1994).
\bibitem{BB}
K.~Abe {\it et al.}  [Belle Collaboration],
Phys.\ Rev.\ D {\bf 66}, 071102 (2002)
[arXiv:hep-ex/0208025];\\
B.~Aubert {\it et al.}  [BABAR Collaboration],
Phys.\ Rev.\ Lett.\  {\bf 89}, 201802 (2002)
[arXiv:hep-ex/0207042].
\bibitem{Gronau}
M.~Gronau and D.~Wyler,
Phys.\ Lett.\ B {\bf 265}, 172 (1991).
\bibitem{Masetti}
M.~Masetti,
Phys.\ Lett.\ B {\bf 286}, 160 (1992).
\bibitem{FW}
R.~Fleischer and D.~Wyler,
Phys.\ Rev.\ D {\bf 62}, 057503 (2000) [arXiv:hep-ph/0004010].
\bibitem{revqq}
V.~V.~Kiselev and A.~K.~Likhoded,
Usp.\ Fiz.\ Nauk\ {\bf 172}, 497 (2002)
[arXiv:hep-ph/0103169];\\
A.~V.~Berezhnoi, V.~V.~Kiselev and A.~K.~Likhoded,
Phys.\ Atom.\ Nucl.\  {\bf 59}, 870 (1996)
[Yad.\ Fiz.\  {\bf 59}, 909 (1996)]
[arXiv:hep-ph/9507242];\\
S.~P.~Baranov,
Phys.\ Rev.\ D {\bf 54}, 3228 (1996);\\
M.~A.~Doncheski, J.~Steegborn and M.~L.~Stong,
Phys.\ Rev.\ D {\bf 53}, 1247 (1996)
[arXiv:hep-ph/9507220];\\
A.~V.~Berezhnoi, V.~V.~Kiselev, A.~K.~Likhoded and A.~I.~Onishchenko,
Phys.\ Rev.\ D {\bf 57}, 4385 (1998)
[arXiv:hep-ph/9710339];\\
V.~V.~Kiselev and A.~E.~Kovalsky,
Phys.\ Atom.\ Nucl.\  {\bf 63}, 1640 (2000)
[Yad.\ Fiz.\  {\bf 63}, 1728 (2000)]
[arXiv:hep-ph/9908321];\\
V.~V.~Braguta, V.~V.~Kiselev and A.~E.~Chalov,
Phys.\ Atom.\ Nucl.\  {\bf 65}, 1537 (2002)
[Yad.\ Fiz.\  {\bf 65}, 1575 (2002)].
\bibitem{QQCP}
V.~V.~Kiselev and O.~P.~Yushchenko,
arXiv:hep-ph/0211382.
\bibitem{Wolf}
L.~Wolfenstein,
Phys.\ Rev.\ Lett.\  {\bf 51}, 1945 (1983).
\bibitem{BBL}
G.~Buchalla, A.~J.~Buras and M.~E.~Lautenbacher,
Rev.\ Mod.\ Phys.\  {\bf 68}, 1125 (1996)
[arXiv:hep-ph/9512380].
\bibitem{QCDSR}
M.~A.~Shifman, A.~I.~Vainshtein and V.~I.~Zakharov,
Nucl.\ Phys.\ B {\bf 147}, 448 (1979),
Nucl.\ Phys.\ B {\bf 147}, 385 (1979),
Nucl.\ Phys.\ B {\bf 147}, 519 (1979);\\
V.~A.~Novikov, L.~B.~Okun, M.~A.~Shifman, A.~I.~Vainshtein,
M.~B.~Voloshin and V.~I.~Zakharov,
Phys.\ Rept.\  {\bf 41}, 1 (1978);\\
L.~J.~Reinders, H.~Rubinstein and S.~Yazaki,
Phys.\ Rept.\  {\bf 127}, 1 (1985).
\bibitem{SR3pt}
P.~Ball, M.~Beneke and V.~M.~Braun,
Nucl.\ Phys.\ B {\bf 452}, 563 (1995) [arXiv:hep-ph/9502300],
Phys.\ Rev.\ D {\bf 52}, 3929 (1995)
[arXiv:hep-ph/9503492];\\
P.~Ball, V.~M.~Braun and H.~G.~Dosch,
Phys.\ Rev.\ D {\bf 48}, 2110 (1993)
[arXiv:hep-ph/9211244];\\
A.~A.~Ovchinnikov and V.~A.~Slobodenyuk,
Sov.\ J.\ Nucl.\ Phys.\  {\bf 50}, 891 (1989) [Yad.\ Fiz.\  {\bf
50}, 1433 (1989)],
Z.\ Phys.\ C {\bf 44}, 433 (1989).
\bibitem{fact}
M.Dugan and B.Grinstein, Phys. Lett. {\bf B255}, 583 (1991);\\
M.A.Shifman, Nucl. Phys. {\bf B388}, 346 (1992);\\
B.Blok, M.Shifman, Nucl. Phys. {\bf B389}, 534 (1993).
\bibitem{KT}
V.V.Kiselev, A.V.Tkabladze, Phys. Rev. {\bf D48}, 5208 (1993).
\bibitem{KLO}
V.V.Kiselev, A.K.Likhoded, A.I.Onishchenko, Nucl. Phys. {\bf
B569}, 473 (2000).
\bibitem{KKL}
V.V.Kiselev, A.K.Likhoded, A.E.Kovalsky, Nucl. Phys. {\bf B585},
353 (2000),
hep-ph/0006104 (2000);
arXiv:hep-ph/0006104.
\bibitem{exBc}
V.~V.~Kiselev,
arXiv:hep-ph/0211021.
\bibitem{HQET}
M.~Neubert,
Phys.\ Rept.\  {\bf 245}, 259 (1994).
\bibitem{NRQCD}
G.T.Bodwin, E.Braaten, G.P.Lepage, Phys. Rev. {\bf D51}, 1125
(1995)
[Erratum-ibid.  {\bf D55}, 5853 (1995)];\\
T.Mannel, G.A.Schuler, Z. Phys. {\bf C67}, 159 (1995).
\bibitem{pNRQCD}
A.~Pineda and J.~Soto,
Nucl.\ Phys.\ Proc.\ Suppl.\  {\bf 64}, 428 (1998)
[arXiv:hep-ph/9707481].
\bibitem{vNRQCD}
M.~E.~Luke, A.~V.~Manohar and I.~Z.~Rothstein,
Phys.\ Rev.\ D {\bf 61}, 074025 (2000) [arXiv:hep-ph/9910209];\\
A.~V.~Manohar and I.~W.~Stewart,
Phys.\ Rev.\ D {\bf 63}, 054004 (2001) [arXiv:hep-ph/0003107].
A.~H.~Hoang, A.~V.~Manohar and I.~W.~Stewart,
Phys.\ Rev.\ D {\bf 64}, 014033 (2001) [arXiv:hep-ph/0102257].
\bibitem{latt}
C.~Bernard {\it et al.}  [MILC Collaboration],
dynamical quarks,'' arXiv:hep-lat/0206016.
\bibitem{Faust'}
D.~Ebert, R.~N.~Faustov and V.~O.~Galkin,
Mod.\ Phys.\ Lett.\ A {\bf 17}, 803 (2002) [arXiv:hep-ph/0204167].
\bibitem{scale}
V.~V.~Kiselev, Phys.\ Part.\ Nucl.\  {\bf 31}, 538 (2000)
[Fiz.\ Elem.\ Chast.\ Atom.\ Yadra {\bf 31}, 1080 (2000)];\\
V.~V.~Kiselev, Int.\ J.\ Mod.\ Phys.\  {\bf A11}, 3689 (1996);\\
V.~V.~Kiselev, Nucl.\ Phys.\ {\bf B406}, 340 (1993).
\bibitem{PMK}
V.~V.~Kiselev,
Mod.\ Phys.\ Lett.\ A {\bf 10}, 1049 (1995)
[arXiv:hep-ph/9409348];\\
V.~V.~Kiselev, Int.\ J.\ Mod.\ Phys. A{\bf 9}, 4987 (1994).
\bibitem{vary}
A.~Abd El-Hady, J.~H.~Munoz and J.~P.~Vary,
Phys.\ Rev.\ D {\bf 62}, 014019 (2000) [arXiv:hep-ph/9909406].
\bibitem{chch}
C.~H.~Chang and Y.~Q.~Chen,
Phys.\ Rev.\ D {\bf 49}, 3399 (1994).
\bibitem{narod}
A.Yu.Anisimov, I.M.Narodetskii, C.Semay, B.Silvestre--Brac, Phys.
Lett. {\bf
B452}, 129 (1999);\\
A.Yu.Anisimov, P.Yu.Kulikov, I.M.Narodetsky, K.A.Ter-Martirosian,
Phys. Atom. Nucl. {\bf 62}, 1739 (1999), Yad. Fiz. {\bf 62}, 1868
(1999).
\bibitem{IKP}
M.~A.~Ivanov, J.~G.~Korner and O.~N.~Pakhomova,
Phys.\ Lett.\ B {\bf 555}, 189 (2003) [arXiv:hep-ph/0212291].
\bibitem{CdF}
P.Colangelo, F.De Fazio, Phys. Rev. {\bf D61}, 034012 (2000).
\bibitem{LHCB}
I.~P.~Gouz, V.~V.~Kiselev, A.~K.~Likhoded, V.~I.~Romanovsky and
O.~P.~Yushchenko,
arXiv:hep-ph/0211432.
\end{thebibliography}
\end{document}